\documentclass[prl,amsmath,twocolumn,showpacs,superscriptaddress]{revtex4}
\usepackage{graphicx}
\usepackage{subfigure}
\newlength{\picwidth}
\begin{document}
%\draft

%%%%%%%%%%%%%%%%%%%%%%%%%%%%%%%%%%%%%%%%%%%%%%%%%%%%%%%%%%%%%%%%%%
\title{Localization in Correlated Bi-Layer Structures:\\
       From Photonic Cristals to Metamaterials and Electron Superlattices}
%%%%%%%%%%%%%%%%%%%%%%%%%%%%%%%%%%%%%%%%%%%%%%%%%%%%%%%%%%%%%%%%%%

%%%%%%%%%%%%%%%%%%%%%%%%%%%%%%%%%%%%%%%%%%%%%%%%%%%%%%%%%%%%%%%%%%
\author{F.~M.~Izrailev}
\affiliation{Instituto de F\'{\i}sica, Universidad Aut\'{o}noma
de Puebla, Apdo. Post. J-48, Puebla 72570, M\'{e}xico}

\affiliation{Department of Physics and Astronomy, Michigan
State University, East Lansing, Michigan 48824-1321, USA}

%%%%%%%%%%%%%%%%%%%%%%%%%%%%%%%%%%%%%%%%%%%%%%%%%%%%%%%%%%%%%%%%%%
\author{N.~M.~Makarov}
\affiliation{Instituto de Ciencias, Universidad Aut\'{o}noma de Puebla,
             Priv. 17 Norte No. 3417, Col. San Miguel Hueyotlipan,
             Puebla 72050, M\'{e}xico.}
%%%%%%%%%%%%%%%%%%%%%%%%%%%%%%%%%%%%%%%%%%%%%%%%%%%%%%%%%%%%%%%%%%

\date{\today}

%%%%%%%%%%%%%%%%%%%%%%%%%%%%%%%%%%%%%%%%%%%%%%%%%%%%%%%%%%%%%%%%%%
\begin{abstract}
In a unified approach, we study the transport properties of
periodic-on-average bi-layered photonic crystals, metamaterials
and electron superlattices. Our consideration is based on the
analytical expression for the localization length derived for
the case of weakly fluctuating widths of layers, that also
takes into account possible correlations in disorder. We
analyze how the correlations lead to anomalous properties of
transport. In particular, we show that for quarter stack
layered media specific correlations can result in a
$\omega^2$-dependence of the Lyapunov exponent in {\it all}
spectral bands.
\end{abstract}
%%%%%%%%%%%%%%%%%%%%%%%%%%%%%%%%%%%%%%%%%%%%%%%%%%%%%%%%%%%%%%%%%%


\maketitle

{\it Introduction.} In recent years much attention was paid to
the propagation of waves (electrons) in periodic
one-dimensional structures (see, e.g. \cite{MS08} and
references therein). The interest to this subject is due to
various applications in which one needs to create materials,
metamaterials or electron superlattices with given transmission
properties. One of the important problems that still remains
open, is the influence of a disorder that can not be avoided in
experimental devices. Such a disorder can be manifested by
fluctuations of the width of layers and distance between
layers, or due to variations of the medium parameters, such as
dielectric constant, magnetic permeability or barrier hight
(for electrons)
\cite{MCMM93,So00,SSS01,Po03,Eo06,No07,Po07,Ao07}.

As is well known, the main quantity that absorbs the influence of a
disorder is the localization length entirely determining transport
properties in a 1D geometry \cite{LGP88}. In contrast to many
studies of the wave (electron) propagation through random
structures, mainly based on various numerical methods, in this
Letter we develop an analytical approach allowing us to derive the
unique expression for the localization length that is valid for
photonic crystals, metamaterials and electron superlattices. Another
key point of consideration is that we explicitly take into account
possible correlations within a disorder, that may be imposed
experimentally. As was recently shown, both theoretically
\cite{IK99,IKU01,IM,HIT08} and experimentally \cite{Bo99,KIKS00},
specific long-range correlations can significantly enhance or
suppress the localization length in desired windows of frequency of
incident waves.

%%%%%%%%%%%%%%%%%%%%%%%%%%%%%%%%%%%%%%%%%%%%%%%%%%%%%%%%%%%%%%%%%%

{\it Model.} We consider the propagation of electromagnetic
wave of the frequency $\omega$ through an infinite array of two
alternating $a$ and $b-$layers (slabs). The slabs are specified
by the dielectric constant $\varepsilon_{a,b}$, magnetic
permeability $\mu_{a,b}$, refractive index
$n_{a,b}=\sqrt{\varepsilon_{a,b}\mu_{a,b}}$, impedance
$Z_{a,b}=\sqrt{\mu_{a,b}/\varepsilon_{a,b}}$ and wave number
$k_{a,b}=\omega n_{a,b}/c$. We assume that the $z$-axis is
directed along the array of bi-layers perpendicular to the
stratification. Within the layers, the electric field obeys the
wave equation,
\begin{equation}\label{WaveEqBC}
\left(\frac{d^2}{dz^2}+k_{a,b}^2\right)\psi_{a,b}(z)=0,
\end{equation}
with two boundary conditions on the interfaces $z=z_{i}$ between
slabs, $\psi_a(z_i)=\psi_b(z_i)$ and
$\mu_a^{-1}\psi'_a(z_i)=\mu_b^{-1}\psi'_b(z_i)$.

A disorder is incorporated in the structure via the random
widths of the slabs, $\widetilde{a}(n)=a+\varrho_a(n)$,
$\widetilde{b}(n)=b+\varrho_b(n)$. Here $n$ enumerates the
elementary $ab$-cells, $a$ and $b$ are the average widths of
layers and $\varrho_a(n)$, $\varrho_b(n)$ stand for small
variations of the widths. In the absence of disorder the array
of slabs is periodic with the period $d=a+b$. The random
sequences $\varrho_{a,b}(n)$ are supposed to be statistically
homogeneous with zero average,
$\langle\varrho_{a,b}(n)\rangle=0$, and the correlations are
fully determined by the binary correlation functions
\begin{eqnarray}\label{Corr-def}
&&\langle\varrho_j(n)\varrho_j(n')\rangle=
\langle\varrho^2_j(n)\rangle K_j(n-n'),\,\,\,\,j=a,b\label{Ka}\\[6pt]
&&\langle\varrho_a(n)\varrho_b(n')\rangle=
\langle\varrho_a(n)\varrho_b(n)\rangle K_{ab}(n-n').\label{Kab}
\end{eqnarray}
In what follows the average $\langle ... \rangle$ is performed over
the whole array of layers or due to the ensemble averaging, that is
assumed to be the same. The two-point auto-correlators
$K_{a,b}(n-n')$ as well as the inter-correlator $K_{ab}(n-n')$ are
normalized to one, $K_{a,b}(0)=K_{ab}(0)=1$. The variances
$\langle\varrho^2_{a,b}(n)\rangle$ are of positive value, while
$\langle\varrho_a(n)\varrho_b(n)\rangle$ can be both positive and
negative. Note that $|\langle\varrho_a(n)\varrho_b(n)\rangle|=
\sqrt{\langle\varrho^2_a(n)\rangle\langle\varrho^2_b(n)\rangle}$. We
assume the positional disorder be weak,
$k^2_{a,b}\langle\varrho^2_{a,b}(n)\rangle\ll1$, allowing us to use
an appropriate perturbation theory. In this case, all transport
properties are entirely determined by the randomness power spectra
$\mathcal{K}_a(k)$, $\mathcal{K}_b(k)$, and $\mathcal{K}_{ab}(k)$,
defined by the relation,
$\mathcal{K}(k)=\sum_{r=-\infty}^{\infty}K(r)\exp(-ikr)$. All the
correlators $K(r)$ are real and even functions of the difference
$r=n-n'$. Because of this fact and due to their positive
normalization, the corresponding Fourier transforms $\mathcal{K}(k)$
are real, even and non-negative functions of the dimensionless
lengthwise wave-number $k$.

%%%%%%%%%%%%%%%%%%%%%%%%%%%%%%%%%%%%%%%%%%%%%%%%%%%%%%%%%%%%%%%%%%

{\it Method.} Our aim is to derive the localization length (LL)
$l_\infty(\omega)$ in the general case of either white or
colored disorder. On the scale of individual slabs the solution
of Eq.~\eqref{WaveEqBC} can be presented in the form of two
maps for $n$th $a$ and $b$ layer, respectively, with
corresponding phase shifts
$\widetilde{\varphi}_{a,b}(n)=\varphi_{a,b}+\xi_{a,b}(n)$,
where $\varphi_{a}=k_{a}a$, $\varphi_{b}=k_{b}b$, and
$\xi_{a,b}(n)=k_{a,b}\varrho_{a,b}(n)$. By combining these maps
with the use of the boundary conditions, one can write the map
for the whole $n$th elementary $ab$-cell,
\begin{equation}\label{map-xy}
x_{n+1}=\widetilde{A}_nx_n+\widetilde{B}_ny_n,\qquad
y_{n+1}=-\widetilde{C}_nx_n+\widetilde{D}_ny_n.
\end{equation}
Here $x_n =\psi_a(z_{an})$ and $y_n= k_a^{-1}\psi'_a(z_{an})$,
the index $n$ corresponds to the left edge while $n+1$ stands
for the right edge of the $n$th cell. The constants
$\widetilde{A}_n$, $\widetilde{B}_n$, $\widetilde{C}_n$,
$\widetilde{D}_n$ depend on $\widetilde{\varphi}_a(n)$,
$\widetilde{\varphi}_b(n)$ and on $Z_a/Z_b=k_b\mu_a/k_a\mu_b$.

Eq.~\eqref{map-xy} can be treated as the map of a linear oscillator
with time-dependent parametric force \cite{IKT95}. Without disorder
the trajectory $x_n,y_n$ creates an ellipse in the phase space
$(x,y)$, that is an image of the unperturbed motion. It is
convenient to make the transformation,
$x_{n}=\upsilon^{-1}Q_n\cos\tau-\upsilon P_n\sin\tau$,
$y_{n}=\upsilon^{-1}Q_n\sin\tau+\upsilon P_n\cos\tau$, to new
coordinates $Q_n,P_n$, in which the unperturbed trajectory occupies
the circle, $Q_{n+1}=Q_n\cos\gamma+P_n\sin\gamma$,
$P_{n+1}=-Q_n\sin\gamma+P_n\cos\gamma$, in the phase space $(Q,P)$.
Here $\tau$ and $\upsilon$ can be found from Eq.~\eqref{map-xy}, and
$\gamma$ determines the Bloch wave number $\kappa=\gamma/d$ arising
in the relation $\psi(z+d)=\exp(i\kappa d)\psi(z)$ for the periodic
array,
\begin{equation}\label{gamma}
\cos\gamma=\cos\varphi_a\cos\varphi_b-\frac{1}{2}
\left(\frac{Z_b}{Z_a}+\frac{Z_a}{Z_b} \right)
\sin\varphi_a\sin\varphi_b .
\end{equation}

In order to take into account the disorder, we expand the constants
$\widetilde{A}_n$, $\widetilde{B}_n$, $\widetilde{C}_n$,
$\widetilde{D}_n$ up to the second order in the perturbation
parameters $\xi_{a}(n)\ll1$ and $\xi_{b}(n)\ll1$. Then, one can
transform the variables $x_n,y_n$ into $Q_n,P_n$. After getting the
perturbed map for $Q_n,P_n$, we pass to action-angle variables
$R_n$, $\theta_n$ via the standard transformations,
$Q_n=R_n\cos\theta_n$, $P_n=R_n\sin\theta_n$. This allows us to
derive the relation between $R_{n+1}$ and $R_n$ keeping linear and
quadratic terms in the perturbation,
\begin{eqnarray}\label{Rn-gen}
&&\frac{R_{n+1}^2}{R_n^2}= 1+\xi_{a}(n)V_{a}(n)+\xi_{b}(n) V_{b}(n)
-\xi_{a}^2(n)-\xi_{b}^2(n)\nonumber\\[6pt]&&+\xi_{a}^2(n)W_{a}(n)
+\xi_{b}^2(n)W_{b}(n)+\xi_{a}(n)\xi_{b}(n)W_{ab}(n) ,
\end{eqnarray}
where $V_{a}(n)$, $V_{b}(n)$, $W_{a}(n)$, $W_{b}(n)$, $W_{ab}(n)$
are complicated functions of $\theta_n$ and the model parameters.

%%%%%%%%%%%%%%%%%%%%%%%%%%%%%%%%%%%%%%%%%%%%%%%%%%%%%%%%%%

{\it Localization length.} The LL can be expressed via the Lyapunov
exponent (LE) $\lambda=d/l_\infty(\omega)$ defined by
\cite{{IKT95}},
\begin{equation}\label{Lyap-def}
\lambda=\frac{1}{2}\langle\ln\left(\frac{R_{n+1}}{R_n}\right)^2\rangle.
\end{equation}
In deriving the LE it was assumed that the distribution of
$\theta_n$ is homogenous within the first order of approximation.
This assumption is correct apart from the band edges,
$\gamma=0;\pi$, and the vicinity of the center, $\gamma=\pi/2$
\cite{HIT08}. The calculation of the LE has been done with the use
of the method developed in Refs.~\cite{IK99,IKU01,HIT08}. Omitting
all details, here we refer to the final result for the total LE,
\begin{eqnarray}\label{LyapFin}
\lambda&=&\frac{\varpi^2}{8\sin^2\gamma}
\Big[\sigma^2_a\mathcal{K}_a(2\gamma)\sin^2\varphi_b+
\sigma^2_b\mathcal{K}_b(2\gamma)\sin^2\varphi_a
\nonumber\\[6pt]
&&-2\sigma^2_{ab}\mathcal{K}_{ab}(2\gamma)
\sin\varphi_a\sin\varphi_b\cos\gamma\Big],
\end{eqnarray}
where $\sigma^2_{a,b}=k_{a,b}^2 \langle\varrho^2_{a,b}(n)\rangle$,
$\sigma^2_{ab}=k_{a}k_{b}\langle \varrho_a(n) \varrho_b(n) \rangle$,
and $\varpi=Z_a/Z_b-Z_b/Z_a$ is the \emph{mismatching factor}. The
expression \eqref{LyapFin} generalizes the results obtained in
Refs.~\cite{BW85,IKU01,HIT08} for particular cases, and is in a
complete correspondence with them. Let us now discuss the derived
expression in some applications.

%%%%%%%%%%%%%%%%%%%%%%%%%%%%%%%%%%%%%%%%%%%%%%%%%%%%%%%%%%%%%%%%%%

{\it Conventional Photonic Layered Media}. In this case all the
optical characteristics, $\varepsilon_{a,b}$, $\mu_{a,b}$,
$n_{a,b}$, $Z_{a,b}$ are \emph{positive constants}. One can see that
if the impedances of $a,b$-slabs are equal, $Z_a=Z_b$, the
mismatching factor $ \varpi$ entering Eq.~\eqref{LyapFin} vanishes
and the perfect transparency emerges ($\lambda=0$) even in the
presence of a disorder. Since the layers are perfectly matched, this
conclusion is general \cite{MS08}, and does not depend on the
strength of disorder. According to Eq.~\eqref{gamma}, in such a case
the stack-structure is effectively equivalent to the homogeneous
medium with the linear spectrum, $\kappa=\omega\overline{n}/c$, that
has no gaps, and where the refractive index is
$\overline{n}=(n_aa+n_bb)/(a+b)$.

For the Fabry-Perot resonances appearing when $\omega/c=s_a\pi/n_aa$
and $\omega/c=s_b\pi/n_bb$, with $s_{a,b}=1,2,3,\dots$, the factors
$\sin\varphi_a$ and $\sin\varphi_b$ in Eq.~\eqref{LyapFin} vanish,
thus giving rise to the \emph{resonance increase} of the LL. In a
special case when $n_aa/n_bb=s_a/s_b$ some resonances from different
layers coincide and would give rise to the divergence of the LL.
However, such a situation can arise only at the edges of spectral
bands, where $\gamma=0,\pi$ and $\sin^2\gamma$ also vanishes.
Therefore, the LL gets a finite value at these points instead of
diverging (due to resonances) or vanishing (due to band edges). The
above statement is not valid at the bottom of the spectrum,
$\omega/c=0$, where the analysis has to be done separately. Note,
nevertheless, that for a white noise,
$\mathcal{K}_{a,b}=1,\,\mathcal{K}_{ab}=0,\pm1$, the LE obeys the
conventional dependence $\lambda\propto\omega^2$ for $\omega\to0$.

A special interest is in the long-range correlations leading to the
divergence of the LL in the controlled windows of frequency
$\omega$. This effect is similar to that found in more simple 1D
models with the correlated disorder \cite{IK99,IKU01,HIT08,KIKS00}.
In our model this effect is due to a possibility to have the
vanishing values of all Fourier transforms,
$\mathcal{K}_{a,b}=\mathcal{K}_{ab}=0$, in some range of frequency
$\omega$. This fact is important in view of experimental
realizations of random disorder with specific correlations. In
particular, one can artificially construct an array of random
bi-layers with such power spectra that abruptly vanishes within
prescribed intervals of $\omega$, resulting in the divergence of the
LL. Also, specific correlations \cite{DIKR08} in a disorder can be
used to ``kill" a sharp frequency dependence associated with the
term $\sin^2\gamma$ in the denominator of Eq.~\eqref{LyapFin}. It is
noteworthy that in the middle of spectral Bloch bands
($\gamma=\pi/2$), the third term vanishes and the inter-correlations
do not contribute to the LL.

The typical dependence $\lambda(\omega)$ for the conventional
photonic bi-layer stack is shown in Fig.~1.

\begin{figure}[h]
\begin{center}
\subfigure{\includegraphics[scale=0.21]{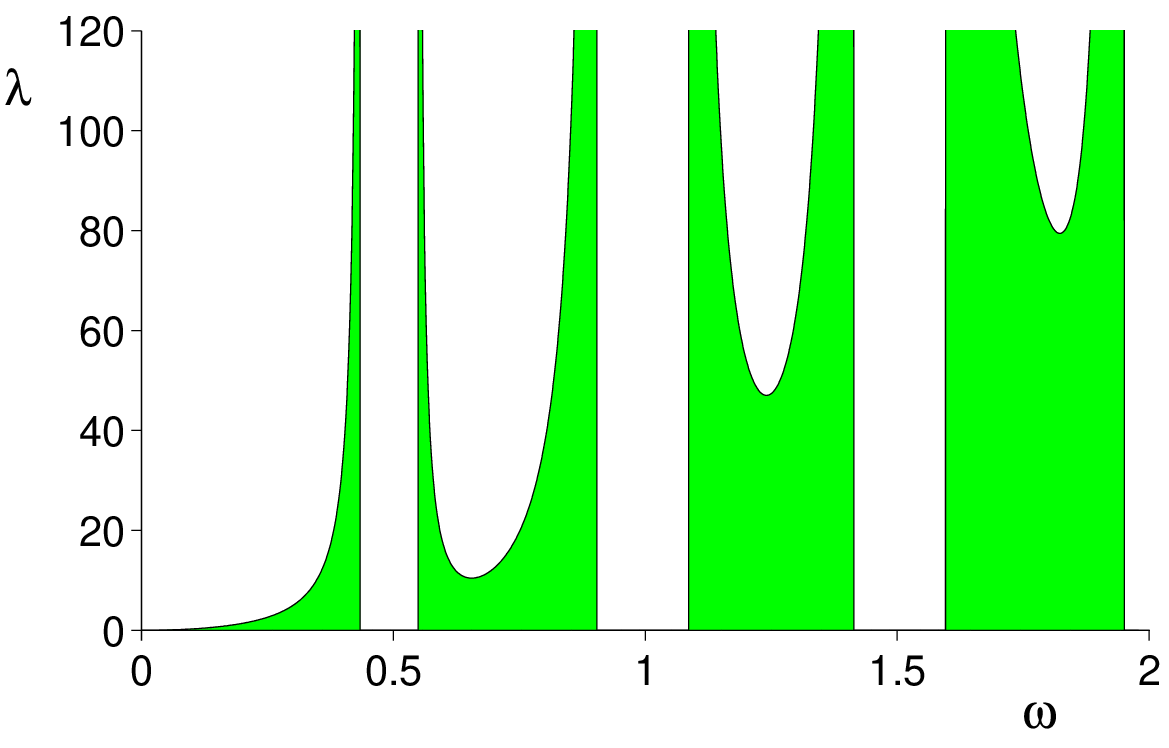}}
\subfigure{\includegraphics[scale=0.21]{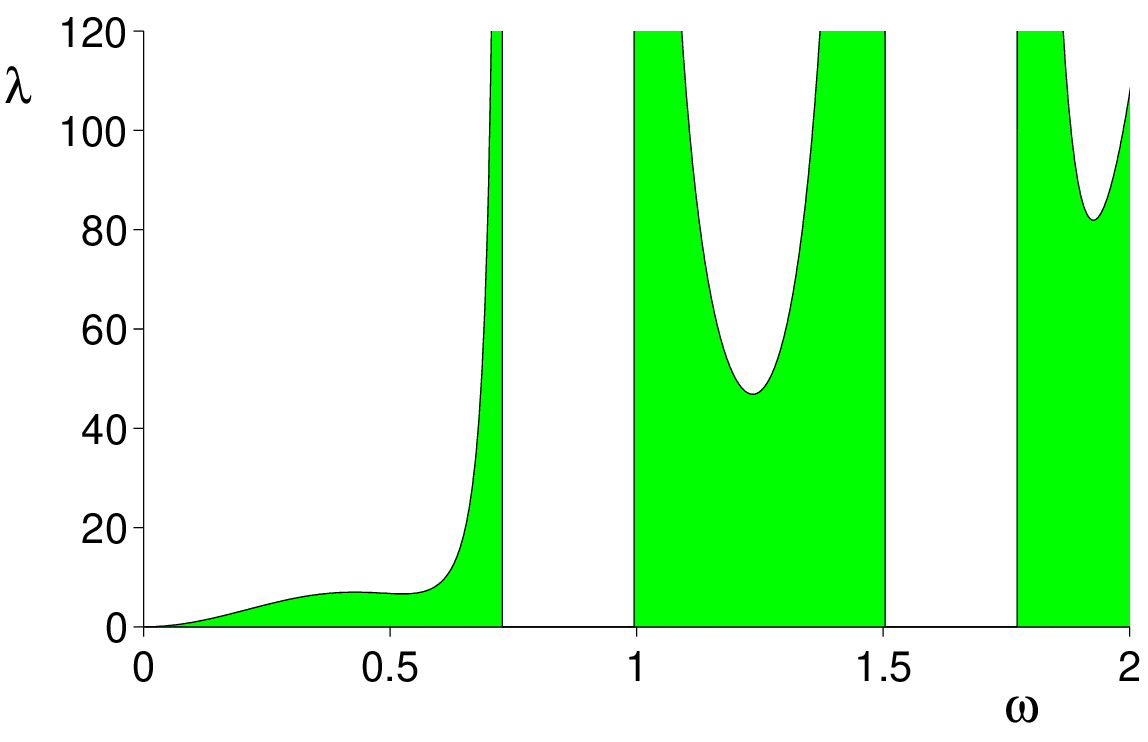}}
\end{center}
\vspace{-0.5cm} \caption{Lyapunov exponent $\lambda$ versus
frequency $\omega$ in arbitrary units for
$\varpi^2n_a^2\langle\varrho^2_{a}(n)\rangle/(2c^2)\approx12.28$,
$\varpi^2n_b^2\langle\varrho^2_{b}(n)\rangle/(2c^2)\approx0.27$ and
$n_aa/c=1.6$, $n_bb/c=0.4$. Left: photonic layered medium,
Eqs.~\eqref{LyapFin} and \eqref{gamma}. Right: the RH-LH bi-layers
with ``plus" in Eq.~\eqref{gamma}.}
\end{figure}

%%%%%%%%%%%%%%%%%%%%%%%%%%%%%%%%%%%%%%%%%%%%%%%%%%%%%%%%%%%%%%%%%%

{\it Quarter Stack Layered Medium.} This term is typically used when
two basic layers, $a$ and $b$, have the same optical width,
$n_aa=n_bb$ (see, e.g. \cite{MS08}). Since $\varphi_a=\varphi_b$, in
this case the dispersion relation \eqref{gamma} takes the form,
\begin{equation}\label{DispRel-QS}
\cos\gamma=1-\frac{1}{2}\frac{(Z_a+Z_b)^2}{Z_aZ_b}\sin^2(k_aa).
\end{equation}
One can see that starting from the second band the top of every even
band coincides with the bottom of next odd band at $\gamma=0$. The
gaps arise only at $\gamma=\pi$.

With the use of Eq.~\eqref{DispRel-QS} the LE can be written as
\begin{eqnarray}\label{Lyap-QS}
\lambda&=&\frac{\mathcal{Z}f(\gamma)}{8\cos^2(\gamma/2)},\qquad
\mathcal{Z}=\frac{(Z_a-Z_b)^2}{Z_a Z_b}, \\[6pt]
f(\gamma)&=&\sigma^2_{a}\mathcal{K}_a(2\gamma)+
\sigma^2_{b}\mathcal{K}_b(2\gamma)-
2\sigma_{ab}^2\mathcal{K}_{ab}(2\gamma)\cos\gamma. \nonumber
\end{eqnarray}
Thus, the LE is finite or vanishes at $\gamma=0$ and diverges at
$\gamma=\pi$.

It is instructive to analyze the simplest case of correlations when
either $\xi_a(n)=\xi_b(n)$ ({\it plus-correlations}) or
$\xi_a(n)=-\xi_b(n)$ ({\it minus-correlations}). In this case one
gets $f(\gamma)=2\sigma^2_{a}\mathcal{K}_a(2\gamma)(1\mp
\cos\gamma)$, correspondingly. As a result, for the LE one can
obtain,
\begin{equation}\label{Lyap-PM}
\lambda_{+}=\frac{\mathcal{Z}\sigma_a^2\mathcal{K}_a(2\gamma)}{2}
\tan^2\frac{\gamma}{2};\qquad
\lambda_{-}=\frac{\mathcal{Z}\sigma_a^2\mathcal{K}_a(2\gamma)}{2}.
\end{equation}
As one can see, $\lambda_{+}\propto\omega^4$ at the bottom of
the spectrum ($\omega\to0$), in contrast to the conventional
dependence $\lambda\propto\omega^2$. Another non-conventional
dependence, $\lambda\propto\omega^6$ was recently found
\cite{Ao07} in a different layered model with left-handed
material. It is interesting that for the minus-correlations and
$\mathcal{K}_a(2\gamma)=1$, the $\omega$-dependence of the LE
is quadratic for any energy inside the spectral bands (see
Fig.~2). In this case the total optical length is constant
within any pair of $a,b$-layers, although the width of both
$a,b$-layers fluctuates randomly. For such correlations, the
quadratic $\omega$-dependence seems to remain within the
non-perturbative regime as well.

\begin{figure}[h]
\begin{center}
\subfigure{\includegraphics[scale=0.27]{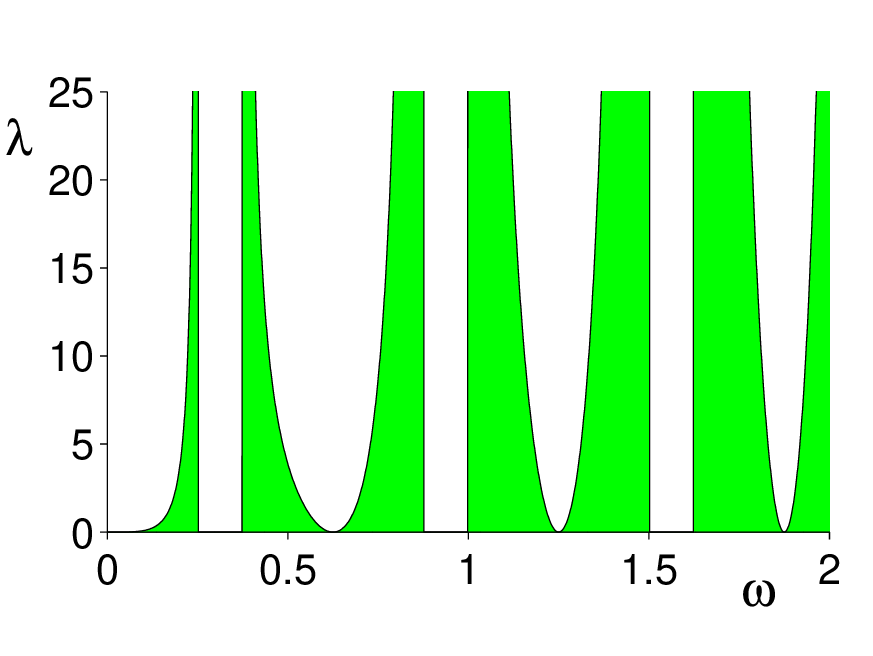}}
\subfigure{\includegraphics[scale=0.27]{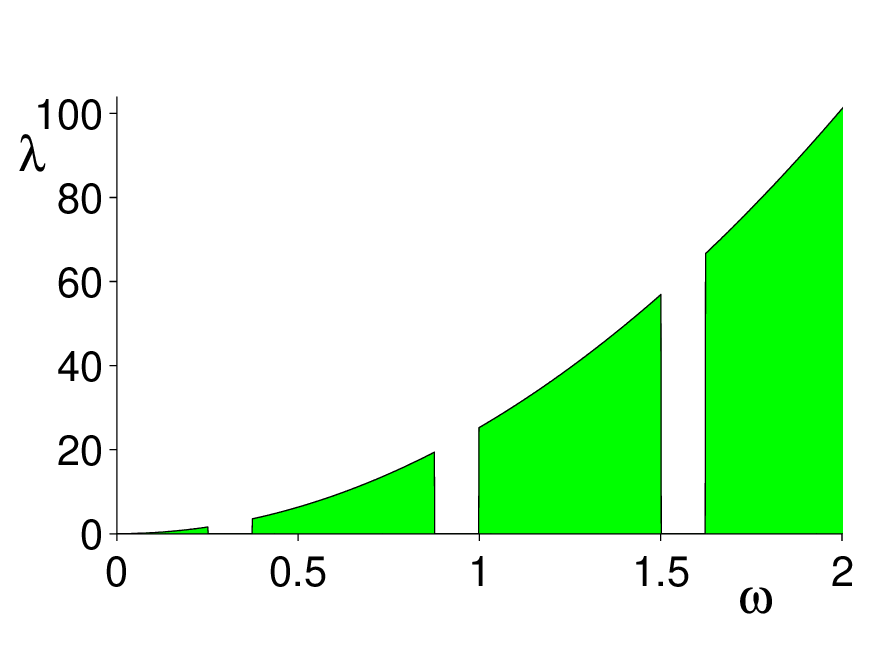}}
\end{center}
\vspace{-0.5cm} \caption{Lyapunov exponent versus frequency for
quarter stack layered medium, for the ``plus correlations" (left)
and ``minus correlations" (right), see Eqs.~\eqref{Lyap-PM} and
\eqref{DispRel-QS}. Here $(Z_a+Z_b)^2/Z_aZ_b=2.4$,
$\mathcal{Z}\langle\varrho^2_{a}(n)\rangle(\pi n_a/c)^2=3.2$,
$\mathcal{K}_a=1$.}
\end{figure}

%%%%%%%%%%%%%%%%%%%%%%%%%%%%%%%%%%%%%%%%%%%%%%%%%%%%%%%%%%%%%%%%%%

{\it Metamaterials.} A special interest is in the mixed system
in which the $a$-layer is a conventional right-handed (RH)
material and $b$-layer is a left-handed (LH) material. This
means that $\varepsilon_a,\mu_a,n_a>0$, whereas
$\varepsilon_b,\mu_b,n_b<0$. However, both impedances remain
positive, $Z_a,Z_b>0$. Remarkably, in comparison with the
conventional stack-structure, the expression \eqref{LyapFin}
for the LE in this case stays the same. The only difference is
that in the dispersion equation \eqref{gamma} the sign ``plus"
has to substitute for ``minus" at the second term (the phase
$\varphi_{b}\equiv k_{b}b=-\omega|n_b|b/c$). Such a ``minor"
correction can drastically change the frequency dependence of
LE, see Fig.~1. Nevertheless, the LE caused by positional
disorder, typically obeys the conventional dependence,
$\lambda\propto\omega^2$ when $\omega\to0$.

Note also that the ideal mixed stack ($\varepsilon_a=\mu_a=n_a=1$,
$\varepsilon_b=\mu_b=n_b=-1$, $Z_a=Z_b=1$) has perfect transmission,
$\lambda=0$, independently of a positional disorder.

One of the interesting features of the mixed layered structures
is that for the RH-LH quarter stack ($n_aa=|n_b|b$) the average
refractive index vanishes,
$\overline{n}=(n_aa-|n_b|b)/(a+b)=0$. As a consequence, it
follows from Eq.~\eqref{gamma} that the spectral bands
disappear and therefore, the transmission is absent, apart from
a discrete set of frequencies where $k_aa=s\pi$ and $\gamma=0$
($s=0,1,2,\ldots$). Evidently Eq.~\eqref{LyapFin} is not valid
in such a situation and an additional analysis has to be done.

It is important that in reality there is a frequency dependence
of the permittivity $\varepsilon_b(\omega)$ and permeability
$\mu_b(\omega)$ \cite{MS08}. This fact is crucial in
applications. In particular, it leads to the following
peculiarities. First, the mismatching factor $\varpi$ in
Eq.~\eqref{LyapFin} can vanish for specific values of frequency
$\omega$ only, thus resulting in a resonance-like dependence
for the transmission. Second, for typical frequency
dependencies the refractive index of $b$-slabs takes an
imaginary value giving rise to the emergence of new gaps.
Specifically, such a gap can arise at the origin of spectrum,
$\omega=0$, in contrast to conventional photonic crystals. It
can be seen that in many aspects the wave transport through the
bi-layered metamaterials is resembling to that of the electrons
through double-barrier structures.

%%%%%%%%%%%%%%%%%%%%%%%%%%%%%%%%%%%%%%%%%%%%%%%%%%%%%%%%%%%%%%%%%%

{\it Electrons.} The developed approach can be also applied to
the propagation of electrons through the bi-layer structures
with alternating potential barriers of the amplitudes $U_a$ and
$U_b$ and slightly perturbed widths. Indeed, the stationary 1D
Schr\"{o}dinger equation for an electron with effective masses
$m_a$, $m_b$ inside the barriers and total energy $E$ can be
written in the form of Eq.~\eqref{WaveEqBC}, in which the
partial wave numbers are associated with the barriers,
$k_a=\sqrt{2m_a(E-U_a)}/\hbar$ and
$k_b=\sqrt{2m_b(E-U_b)}/\hbar$. Another change, $\mu_{a,b}\to
m_{a,b}$, should be done in the boundary condition on
hetero-interfaces, $m_a^{-1}\psi'_a(z_i)=m_b^{-1}\psi'_b(z_i)$.
Correspondingly, $Z_b/Z_a=k_am_b/k_bm_a$ in the dispersion
relation \eqref{gamma} and in the expression \eqref{LyapFin}
for the LE.

If the energy $E$ is smaller then the heights of both barriers,
$E<U_a,\,U_b$, the electron wave numbers are purely imaginary.
As a consequence, the electron states are strongly localized
and the structure is non-transparent.

For other case when $U_a<E<U_b$, the \emph{tunneling propagation} of
electrons emerges. In this case the wave number $k_a$ is real while
$k_b$ is imaginary. Therefore, the electron moves freely within any
$a$-barrier and tunnels through the $b$-barriers. Thus, the
expressions \eqref{LyapFin} for $\lambda$ and \eqref{gamma} for
$\gamma$ have to be modified according to the change, $k_b \to
i|k_b|$ and $\sin(k_bb) \to i\sinh(|k_b|b)$. As a result, the
Fabry-Perot resonance increase of the LL arises only due to the
second and third terms of Eq.~\eqref{LyapFin}. The general
expression for the LE can be essentially simplified for a particular
case of an array with delta-like potential barriers and slightly
disordered distance between them. The corresponding expression for
the LE is in full correspondence with that obtained in
Ref.~\cite{IKU01}.

\begin{figure}[h]
\vspace{-0.2cm}
\begin{center}
\includegraphics[width=4.5cm,height=3cm]{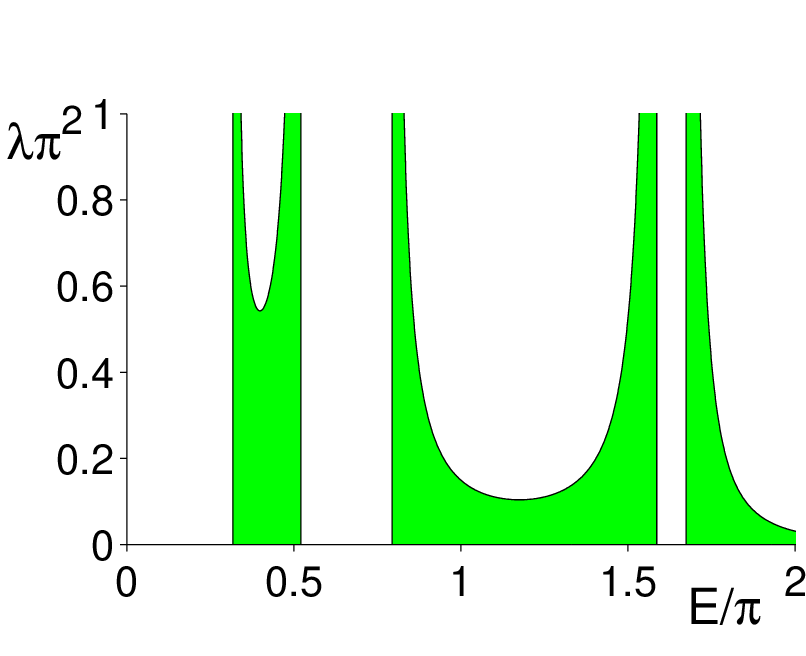}
\end{center}
\vspace{-0.5cm} \caption{ Lyapunov exponent vs. energy for electrons
in bi-layered structure. Here $2m_a/\hbar^2 =2m_b/\hbar^2=1$,
$U_a=0.15$, $U_b=1.6$, $a=0.35\pi$, $b=0.65\pi$,
$\langle\varrho^2_{a}(n)\rangle=(0.1a)^2$,
$\langle\varrho^2_{b}(n)\rangle=(0.2b)^2$ The transition between
tunneling and over-barrier scattering occurs at $E/\pi=0.5$.}
\end{figure}

For the \emph{over-barrier scattering}, when $U_a<U_b<E$, both
wave numbers, $k_a$ and $k_b$, are positive and the electron
transport is similar to that for the conventional photonic
stack but with dispersive parameters. The example of the energy
dependence of the LE is given in Fig.3. It is interesting that
if $E=(U_bm_a-U_am_b)/(m_a-m_b)$, an electron does not change
its velocity in the barriers, $\hbar k_a/m_a=\hbar k_b/m_b$,
although its momentum changes, $\hbar k_a>\hbar k_b$. The LE
vanishes in this case. Thus, such a ``free" electron motion is
equivalent to that in a homogeneous medium with perfect
transmission.

%%%%%%%%%%%%%%%%%%%%%%%%%%%%%%%%%%%%%%%%%%%%%%%%%%%%%%%%%%%%%%%%%%

{\it Conclusion.} We derived the expression for the inverse
localization length for quasi-periodic bi-layer structures, whose
widths are weakly perturbed. Our result can be applied both to
conventional photonic crystals and to metamaterials, as well as to
the electron superlattices. Another feature of the approach is that
it takes into account possible correlations in a disorder that can
lead to anomalous frequency (energy) dependence of transport
properties. Due to the correlations one can significantly enhance or
suppress the transmission/reflection through the bi-layered devices
within the prescribed windows of frequency (energy) of
electromagnetic (electron) waves. Our results may have a strong
impact for the fabrication of a new class of disordered optic
crystals, left/right handed metamaterials, and electron nanodevices
with \emph{selective} transmission and/or reflection.

F.M.I. acknowledges partial support by the VIEP BUAP-2008
grant.

%%%%%%%%%%%%%%%%%%%%%%%%%%%%%%%%%%%%%%%%%%%%%%%%%%%%%%%%%%%%%%%%%%

\end{document}